\documentclass[prd,aps,preprint,tightenlines,showpacs,nofootinbib,superscriptaddress]{revtex4-1}
\usepackage{mathrsfs}
\usepackage{amsfonts}
\usepackage{amsmath}
\usepackage{array}
\usepackage{verbatim}
\usepackage{bm}
\usepackage{epsfig}
\usepackage{graphicx}
\usepackage{relsize}
\RequirePackage{xspace}

\begin{document}
\title{Incoherent vector mesons production in PbPb ultraperipheral collisions at the LHC }
\author{Ya-Ping Xie}\email{xieyaping@impcas.ac.cn}
\affiliation{Institute of Modern Physics, Chinese Academy of
Sciences, Lanzhou 730000, China}
\affiliation{School of Physical Science and Technology, Lanzhou University, Lanzhou 730000, Chnia}
\affiliation{ University of Chinese Academy of Sciences, Beijing 100049, China}
\author{Xurong Chen}\email{xchen@impcas.ac.cn}
\affiliation{Institute of Modern Physics, Chinese Academy of
Sciences, Lanzhou 730000, China}
\begin{abstract}
The incoherent rapidity distributions of vector mesons are computed in dipole model in PbPb ultraperipheral collisions at the CERN Large Hadron Collider (LHC). The IIM model fitted from newer data is employed in the dipole amplitude. The Boosted Gaussian and Gaus-LC wave functions for vector mesons are implemented in the calculation as well. Predictions for the $J/\psi$, $\psi(2s)$, $\rho$ and $\phi$ incoherent rapidity distributions are evaluated and compared with experimental data and other theoretical predictions in this paper. We obtain closer predictions of the incoherent rapidity distributions for $J/\psi$ than previous calculations in the IIM model.
\end{abstract}
\maketitle
\section{introduction}
\indent  Ultrapheripheral collisions (UPCs) at Relativistic Hadron Ion Collider (RHIC) and the LHC provide variable tools to probe structure of the hadrons~\cite{Bertulani:2005ru,Baltz:2007kq}. In recent years, development of experimental technologies helps us measure the rapidity distributions of the vector mesons in PbPb and AuAu UPCs. Some precise experimental data has been listed in recent literatures~\cite{Aaij:2013jxj,Adam:2015sia,Abbas:2013oua,Adam:2015gsa,Abelev:2012ba,TheALICE:2014dwa}. These experimental measurements require theoretical calculations at high energy level. Some theoretical groups have evaluated the predictions for vector mesons production in PbPb and AuAu UPCs in different approaches~\cite{Klein:1999qj,Frankfurt:2002sv,Goncalves:2005yr,Cisek:2012yt,Adeluyi:2012ph,Rebyakova:2011vf,Andrade-II:2015kaa,Lappi:2013am,Lappi:2010dd,Toll:2012mb}.\\
\indent In UPCs, the impact parameter of two nuclei is larger than the sum of two nuclear radius. In  nucleus-nucleus collisions, the nuclei can remain intact or break up. If the two nuclei remain intact, it is called coherent process. If one nucleus or two nuclei break up, it is called incoherent process. The incoherent rapidity distribution of $J/\psi$ has been computed in PbPb and p-Pb UPCs In Ref.~\cite{Lappi:2013am}. The incoherent cross section of light vector meson $\rho$ is evaluated in $\text{eA}$ collisions in Ref.~\cite{Goncalves:2015poa}. In this paper, we will compute the incoherent rapidity distributions for vector mesons in PbPb UPCs at the LHC and give some theoretical predictions for future experiments. \\
\indent In UPCs, the two nuclei interact with each other by photon emission from nucleus. Photon-nucleus interaction has been studied in Deep Inelastic Scattering (DIS) at HERA. In small-$x$ physics, photon-nucleus interaction is described in the dipole model successfully, Dipole amplitude and light-front wave functions are put together in the calculation for the exclusive vector meson cross section in the dipole model. The dipole amplitude contains all QCD dynamics of dipole-proton interaction at high energy, and it is solution of the BK evolution equation~\cite{Balitsky:1995ub}. Impact parameter independent dipole amplitude has been investigated by solving the BK evolution equation in the literatures~\cite{Albacete:2007yr}. But it is difficult to obtain impact parameter dependent dipole amplitude by solving the BK evolution equation. There are various phenomenological models to describe the dipole amplitude. The GBW model was firstly proposed in 1999~\cite{GolecBiernat:1998js}. But it doesn't match the DGLAP evolution equation. Then, the BGBK model was proposed in 2002~\cite{Bartels:2002cj}, which is an extension of the GBW model. Beside, the IIM model was also proposed based on the BK evolution equation in 2004~\cite{Iancu:2003ge,Soyez:2007kg,Ahmady:2016ujw}. The impact parameter dependent BGBK (IP-Sat) and IIM (bCGC) models were proposed in Refs~\cite{Kowalski:2003hm,Kowalski:2006hc,Rezaeian:2012ji,Watt:2007nr,Rezaeian:2013tka}. All these phenomenological models are fitted from proton structure function F2 or reduced cross section from the ZEUS and H1 collaboration. The latest combined ZEUS and H1 data on inclusive DIS is published in Ref.~\cite{Abramowicz:2015mha}. Then, the IIM model is fitted with new parameters in Ref~\cite{Ahmady:2016ujw}. In this paper, we will use newer fit of the IIM model to predict the incoherent rapidity distributions for vector mesons in PbPb UPCs at the LHC. \\
\indent Light-front wave functions of photon and vector meson are also included in the dipole amplitude.  The light-front wave functions of photon can be computed in QED analytically. The light-front wave function of vector meson can not be computed analytically. Various models are used to describe the wave function of vector meson. Some models are constructed as photon structure, for example, the Boosted Gaussian, Gaus-LC and DGKP model~\cite{Kowalski:2006hc,Rezaeian:2012ji,Nemchik:1996cw,Dosch:1996ss,Forshaw:2003ki,Goncalves:2004bp,Xie:2016ino,Forshaw:2012im}.\\
\indent In this paper, we employ the IIM model and use different models of wave functions to predict rapidity distributions of vector mesons in PbPb UPCs at $\sqrt{s_{NN}}=2.76~\text{TeV}$ and $\sqrt{s_{NN}}=5.02~\text{TeV}$. The parameters of the IIM model are fitted from the combined ZEUS and H1 data published in 2015. The goal of this paper is to predict the incoherent rapidity distributions of vector mesons in PbPb UPCs at $\sqrt{s_{\text{NN}}}=2.76~\text{TeV}$ and $\sqrt{s_{\text{NN}}}=5.02~\text{TeV}$ for future experimental plan. This paper is organized as follows. Section II introduces dipole model and distinct approached for the vector meson wave functions. Numerical results are presented in section III. And conclusion is given in section IV.

\section{The dipole model}
In UPCs, the impact parameter is larger than twice radius of the nucleus. Strong interaction is suppressed since two nuclei can not touch each other. Photon can emit from nucleus at high energy. Then, the photon can scatter off the nucleus, and vector meson are produced after photon-nucleus scattering. The vector meson's rapidity distributions in UPCs can be computed as \cite{Bertulani:2005ru}
\begin{eqnarray}
\frac{d\sigma^{A_1A_2}}{dy}=\bigg[n^{A_1}(\omega)\sigma^{\gamma A_2}(\omega)\bigg]_{\omega_{\text{left}}}+\bigg[n^{A_2}(\omega)\sigma^{\gamma A_1}(\omega)\bigg]_{\omega_{\text{right}}}.
\label{dndk}
\end{eqnarray}
Here  $\text{y}$ is the rapidity of the vector meson. $\sigma^{\gamma A}(\omega)$ is the cross section of the photon-nucleus scattering. $n(\omega)$ is the equivalent photon flux in the nucleus, with $\omega_{\text{left}}=\frac{M_V}{2}\exp(-\text{y})$, and $\omega_{\text{right}}=\frac{M_V}{2}\exp(\text{y})$. $M_V$ is the mass of the vector meson. In  nucleus-nucleus scattering, the equivalent photon flux is computed as follow~\cite{Klein:1999qj}
\begin{eqnarray}
n(\omega)=\frac{2 Z^2\alpha_{em}}{\pi}\big[\xi K_1(\xi)
K_0(\xi)-\frac{\xi^2}{2}[K_1^2(\xi)-K_0^2(\xi)]\big],
\end{eqnarray}
   where $Z$ is charge of the nucleus and $\xi=2\omega R_A/\gamma_L$, $R_A$ is the radius of the nuclues, $\gamma_L$ is the lorentz factor, with $\gamma_L=\sqrt{s_{NN}}/2m_p$. $K_0(x)$ and $K_1(x)$ are modified Bessel functions.\\
   \indent $\sigma(\omega)$ is obtained by integrating $|t|$ from differential cross section. In photon-proton scattering, the differential photon-proton cross section is calculated as 
\begin{eqnarray}
\frac{d\sigma^{\gamma p\to Vp}}{dt}=\frac{R_g^2(1+\beta^2)}{16\pi}
\left|\mathcal{A}_{T}(x,\Delta)\right|^2,
\label{dsigma1}
\end{eqnarray}
where $T$ denotes the transverse amplitudes, since the photon is a real photon. $
\beta$ is the ratio of the real amplitude, and $R_g$ accounts the skewness effect~\cite{Shuvaev:1999ce}. The amplitude of $\gamma p\to Vp$ is~\cite{Kowalski:2006hc}
\begin{eqnarray}
\mathcal{A}_{T}(x,\Delta)= i\int d^2\bm r\int_0^1\frac{dz}{4\pi}
\int d^2\bm{b}(\Psi_V^*\Psi_{\gamma})_{T}(z,\bm r)e^{-i\bm{b}\cdot\bm{\Delta} }\frac{d\sigma_{q\bar{q}}}{d^2\bm b}(x,\bm r,\bm b).
\end{eqnarray}
The impact parameter dependent dipole cross section can be rewritten as 
  \begin{eqnarray}
\frac{d\sigma_{q\bar{q}}}{d^2\bm b}(x,\bm r,\bm b)=2T_p(\bm b)\mathcal{N}(x,\bm r),
  \end{eqnarray}
  where $T_p(\bm b)=\exp(-\bm b^2/2B_p)$ is a Gaussian profile function of proton, and $\mathcal{N}(x,r)$ is the amplitude of an elastic dipole-proton scattering amplitude.\\
\indent In photon-nucleus scattering, coherent differential cross section is obtained by squaring the average of amplitude $|\langle \mathcal{A}(x,\Delta)\rangle_N|^2$. The sum of coherent and incoherent differential cross section is obtained by averaging the square of amplitude $\langle\mathcal{A}^2(x,\Delta)\rangle_N$. Then, the incoherent differential cross section can be obtained from the variance of amplitude $\langle\mathcal{A}^2(x,\Delta)\rangle_N-|\langle \mathcal{A}(x,\Delta)\rangle_N|^2$. Since the square of average amplitude is very small at large $\Delta$. Thus, the incoherent differential cross section can be obtained by averaging the square of amplitude  $\langle\mathcal{A}^2(x,\Delta)\rangle_N$ at large $\Delta$. The coherent differential cross section is 
\begin{eqnarray}
\frac{d\sigma^{\text{coh}}}{dt}=\frac{R_g^2(1+\beta^2)}{16\pi}
\left|\langle\mathcal{A}(x,\Delta)\rangle_N\right|^2,\label{dsigma}
\end{eqnarray}
where the average amplitude is calculated as~\cite{Kowalski:2003hm,Lappi:2010dd,Lappi:2013am}
\begin{eqnarray}
\langle\mathcal{A}(x,\Delta)\rangle_N&=& i\int d^2\bm{r}\int_0^1\frac{dz}{4\pi}
\int d^2\bm{b}(\Psi_V^*\Psi_{\gamma})_{T}(z,r)e^{-i\bm{b}\cdot \bm{\Delta} }\notag\\
&&\times2(1-\exp(-2\pi B_pAT_A(\bm b)\mathcal{N}(x,\bm r)).
\label{amp}
\end{eqnarray}
where $T_A(\bm b)$ is the profile function of the nucleus.\\ 
\indent As discussed before, the incoherent differential cross section of $\gamma A\to VX$ in photon-nucleus scattering at large transfer momentum is computed by averaging the square of amplitude at large transfer momentum. It is written as
\begin{eqnarray}
\frac{d\sigma^{\text{incoh}}}{dt}=\frac{R_g^2(1+\beta^2)}{16\pi}\langle\left|\mathcal{A}(x,\Delta)\right|^2\rangle_N,\label{dincoh}
\end{eqnarray}
where the average of square amplitude is calculated as~\cite{Lappi:2010dd,Lappi:2013am,Goncalves:2015poa}
\begin{eqnarray}
\langle\left|\mathcal{A}(x_p,\Delta)\right|^2\rangle_N&=& 16\pi B_pA\int d^2\bm b \int d^2\bm{r}\int d^2\bm{r}^\prime\int_0^1\frac{dz}{4\pi}\int_0^1\frac{dz^\prime}{4\pi}
 [\Psi_V^*\Psi_{\gamma}]_{T}(z,r)[\Psi_V^*\Psi_{\gamma}]_{T}(z^\prime,r^\prime)
 \notag\\&&\times e^{-B_p\Delta^2}e^{-2\pi B_pAT_A(\bm b)[\mathcal{N}(x,r)+\mathcal{N}(x,r^\prime)]}\Bigg(\frac{\pi B_p \mathcal{N}(x,r)\mathcal{N}(x,r^\prime)T_A(\bm b)}{1-2\pi B_pT_A(\bm  b)[\mathcal{N}(x,r)+\mathcal{N}(x,r^\prime)]}\Bigg) \notag\\&\approx&
 16\pi^2B_p^2\int d^2\bm  b\int d^2\bm{r}\int d^2\bm{r}^\prime\int_0^1\frac{dz}{4\pi}\int_0^1\frac{dz^\prime}{4\pi}
  [\Psi_V^*\Psi_{\gamma}]_{T}(z,r)[\Psi_V^*\Psi_{\gamma}]_{T}(z^\prime,r^\prime)\notag\\ &\times& e^{-B_p\Delta^2}\mathcal{N}(x,r)\mathcal{N}(x,r^\prime)AT_A(\bm b)e^{-2\pi(A-1)B_pT_A(\bm b)
  [\mathcal{N}(x,r)+\mathcal{N}(x,r^\prime)]}.
\label{incohamp}
\end{eqnarray}
A similar expression of the incoherent differential cross section can be found in Refs.~\cite{Kopeliovich:2001xj,Goncalves:2009za,Ducati:2013bya}.\\
\indent In Ref.~\cite{Iancu:2003ge,Soyez:2007kg,Ahmady:2016ujw} the dipole cross section is impact parameter independent $\sigma(x,r)=\sigma_0\mathcal{N}(x,r)$. In diffractive process, vector mesons production decreases exponentially with $|t|$ as $e^{-B_p |t|}$. Thus, we can introduce a Gaussian profile function  of impact parameter like $T_p(b)=\exp(-\bm{b^2}/2B_p)$. After integrating impact parameter, we obtain $\sigma_0=4\pi B_p$. Thus, we modified $\sigma(x,r)$ as impact parameter dependent as the same as Ref.~\cite{Marquet:2007nf} $\sigma(x,r,b)=\exp(-\bm b^2/2B_p)\mathcal{N}(x,r).$ 
 The amplitude $\mathcal{N}(x,r)$ in the IIM model is written as ~\cite{Iancu:2003ge,Soyez:2007kg}
 \begin{eqnarray}
 \mathcal{N}(x,r)=\begin{cases}
 \mathcal{N}_0(\frac{rQ_{s}}{2})^{2(\gamma_s+(1/\kappa\lambda Y)\ln(2/rQ_s))},\quad rQ_s\le2,\\
 1-\exp\big(-a\ln^2(b rQ_s)\big),\quad\quad\quad\!\!\!\! rQ_s>2.
 \end{cases}
 \end{eqnarray}
With $Y=\ln(1/x)$, $\mathcal{N}_0=0.7$ and $\kappa=9.9$ ( leading order BFKL prediction), and $Q_s(x)=(x_0/x)^{\lambda/2}$~GeV. \\
 \indent In UPCs, the photon is a real one. Thus, we just consider the transverse overlap of the photon and vector meson.
  We use the same transverse overlap between the photon and vector meson as follow ~\cite{Kowalski:2006hc}
 \begin{eqnarray}
 (\Psi_V^*\Psi_{\gamma})_T(r,z)=e_fe\frac{N_c}{\pi z(1-z)}\lbrace  m_f^2
 K_0(\epsilon r)\phi_T(r,z)-(z^2+(1-z)^2)\epsilon K_1(\epsilon r)\partial_r
 \phi_T(r,z)\rbrace,\notag\\
 \end{eqnarray}
	 where $e=\sqrt{4\pi\alpha_{em}}$,  $m_f$ is the mass of quarks. $e_f$ is the electric charge of  quarks, and with $\epsilon=\sqrt{z(	-z)Q^2+m_f^2}$, in the following calculation, we set $Q^2=0  \text{GeV}^2$.  $N_c$ is the number of colors. The transverse scalar function  $\phi_T(r,z)$  of the Gaus-LC model for ground state of vector meson is given by~\cite{Kowalski:2006hc}
 \begin{eqnarray}
 \phi_T(r,z)=N_T(z(1-z))^2\exp(-\frac{r^2}{2R_T^2}).
 \end{eqnarray}
 The Boosted Gaussian model is a phenomenological model, which is originated from NNPZ model~\cite{Nemchik:1996cw,Forshaw:2003ki}. The scalar function of the Boosted Gaussian model for ground state of vector meson is given by
 \begin{eqnarray}
 \phi^{1s}_T(r,z)=N_Tz(1-z)\exp\big(-\frac{m_f^2\mathcal{R}_{1s}^2}{8z(1-z)}-
 \frac{2z(1-z)r^2}{\mathcal{R}^2_{1s}}+\frac{m_f^2\mathcal{R}^2_{1s}}{2}\big).
 \end{eqnarray}
 The Boosted Gaussian scalar function for $\psi(2s)$ is given by~\cite{Armesto:2014sma}
  \begin{eqnarray}
  \phi^{2s}_T(r,z)&=&N_Tz(1-z)\exp\big(-\frac{m_f^2\mathcal{R}^2_{2s}}{8z(1-z)}-
  \frac{2z(1-z)r^2}{\mathcal{R}^2_{2s}}+\frac{m_f^2\mathcal{R}^2_{2s}}{2}\big)\notag\\
  &\times&\Bigg[1+\alpha_{2s}\Bigg(2+\frac{m_f^2\mathcal{R}^2_{2s}}{8z(1-z)}-
  \frac{4z(1-z)r^2}{\mathcal{R}_{2s}^2}-m_f^2\mathcal{R}_{2s}^2\Bigg)\Bigg].
  \end{eqnarray}
\indent With help of above formulas, we can compute the differential cross section of photon-nucleus scattering. Then, after integrating $|t|$, we obtain the total cross section of photon-nucleus scattering. Eq.~(\ref{dincoh}) is not valid at small $|t|$ with $|t|_\text{min}$ in Eq.~(\ref{dincoh}) being the value of $|t|$ where incoherent and coherent cross sections are equal.  Finally, we can obtain incoherent rapidity distributions of vector mesons in ultraperipheral collisions. 
 \section{Numerical results}
 \indent In this section, we present the numerical results of the incoherent rapidity distributions of $J/\psi$, $\psi(2s)$, $\rho$ and $\phi$ mesons in PbPb UPCs at the LHC. The center energy is $\sqrt{s_\text{NN}}=2.76~\text{TeV}$ and $\sqrt{s_\text{NN}}=5.02~\text{TeV}$.  The $|t_\text{min}|$ of integration of Eq.~(\ref{dincoh}) is $|t|_\text{min}=0.035~\text{GeV}^2$ in the integration of $|t|$. The parameters of the IIM model we used are listed in Table.~\ref{table01}. The parameter $B_p$ in Table.~\ref{table01} is computed from $\sigma_0=29.9~\text{mb}$.
  \begin{table}[htbp]
   \begin{tabular}{p{1.5cm}p{1.5cm}p{1.5cm}p{2.cm}p{1.cm}p{1.cm}p{2cm}p{2cm}}
   \hline
   \hline
   $m_{u,d,s}$& $m_c$ & $\sigma_0$&$B_p$ & $\gamma_s$ & $\lambda$& $x_0$&$ \chi^2/\text{d.o.f}$\\
   \hline
   $0.14$~GeV& 1.27~GeV  &  29.9~mb& 6.12~$\text{GeV}^{-2}$    &0.724 &0.206&$ 6.33\times10^{-6}$& 554/520=1.07 \\
   \hline
   \hline
   \end{tabular}
   \caption{Parameters of the IIM model~\cite{Ahmady:2016ujw}.}
   \label{table01}
   \end{table} 
  For the vector meson wave functions, we use the Boosted Gaussian and Gaus-LC wave functions in our calculations. The parameters of the Boosted Gaussian and Gaus-LC wave functions are listed in Table.~\ref{table02} and Table.~\ref{table03}. 
      \begin{table}[htbp]
      \begin{tabular}{p{1cm} | p{1.5cm}p{1.5cm}p{1.5cm}p{1.5cm}p{1.5cm}p{1.5cm} p{1.5cm}}
      \hline
      \hline
      meson &$e_f$& mass& $f_V$ & $m_f$& $N_T$& $\mathcal{R}^2$&$\alpha_{2s}$ \\
      \hline
      & & GeV & GeV & GeV &   & $\text{GeV}^2 $ & \\
      \hline
      $J/\psi$  &$2/3$      & 3.097    & 0.274  & 1.27      & 0.596      &2.45 & \\
       $\psi(2s)$  &$2/3$      & 3.686  & 0.198  & 1.27      & 0.70      &3.72 & -0.61\\
      $\phi$     &$1/3$     & 1.019    & 0.076  &  0.14    & 0.919     & 11.2   & \\
        $\rho$       &$1/\sqrt{2}$  &   0.776 & 0.156 & 0.14 	  & 0.911	 	& 12.9& \\
      \hline
      \hline
      \end{tabular}
      \caption{Parameters of the Boosted Gaussian wave function for vector mesons~\cite{Kowalski:2006hc,Armesto:2014sma}.}
      \label{table02}
      \end{table} 
    \begin{table}[htbp]
    \begin{tabular}{p{1cm} | p{1.5cm}p{1.5cm}p{1.5cm}p{1.5cm}p{1.5cm}p{1.5cm} }
    \hline
    \hline
    meson &$e_f$& mass& $f_V$ & $m_f$& $N_T$& $R_T^2$ \\
    \hline
    & & GeV & GeV & GeV &   & $\text{GeV}^2 $ \\
    \hline
    $J/\psi$  &$2/3$      & 3.097    & 0.274  & 1.27      & 1.45       & 5.57 \\
    $\phi$     &$1/3$     & 1.019    & 0.076  &  0.14    & 4.75       & 16.9   \\
      $\rho$       &$1/\sqrt{2}$  &   0.776 & 0.156 & 0.14 	  & 4.47	 	& 21.9  \\
    \hline
    \hline
    \end{tabular}
    \caption{Parameters of the Gaus-LC wave function for vector mesons~\cite{Kowalski:2006hc}.}
    \label{table03}
    \end{table} 
    The quark mass is important in the dipole model, which is fitted from inclusive cross section of photon-proton scattering. In Refs.~\cite{Rezaeian:2012ji,Rezaeian:2013tka}, the light quark mass is  $m_{u,d,s}=0.001$~GeV. In the new fit, the light quark mass is $m_{u,d,s}=$0.14 GeV. \\
    \begin{figure}[htbp]
    \centering
       \includegraphics[width=3in]{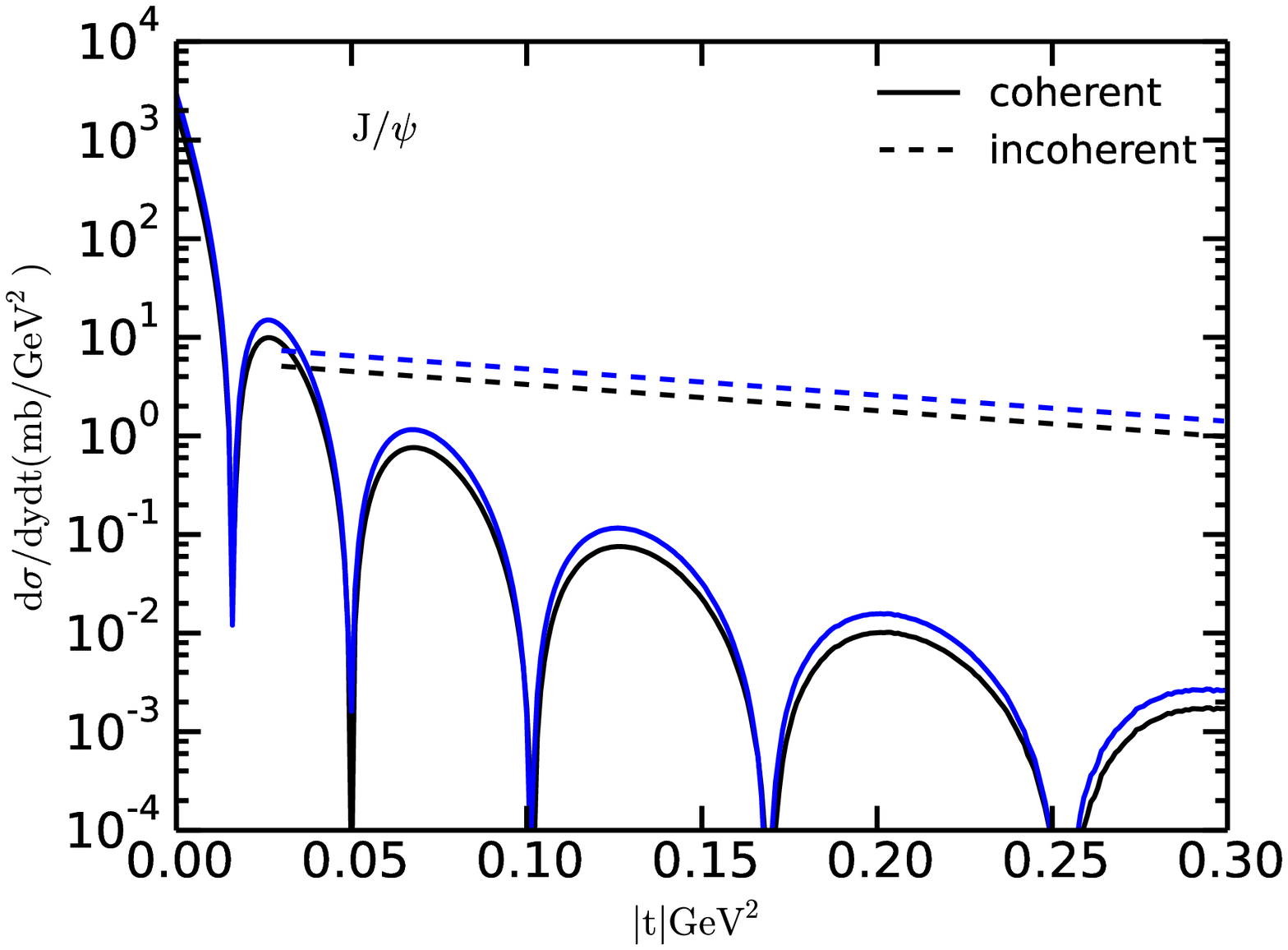}
       \includegraphics[width=3in]{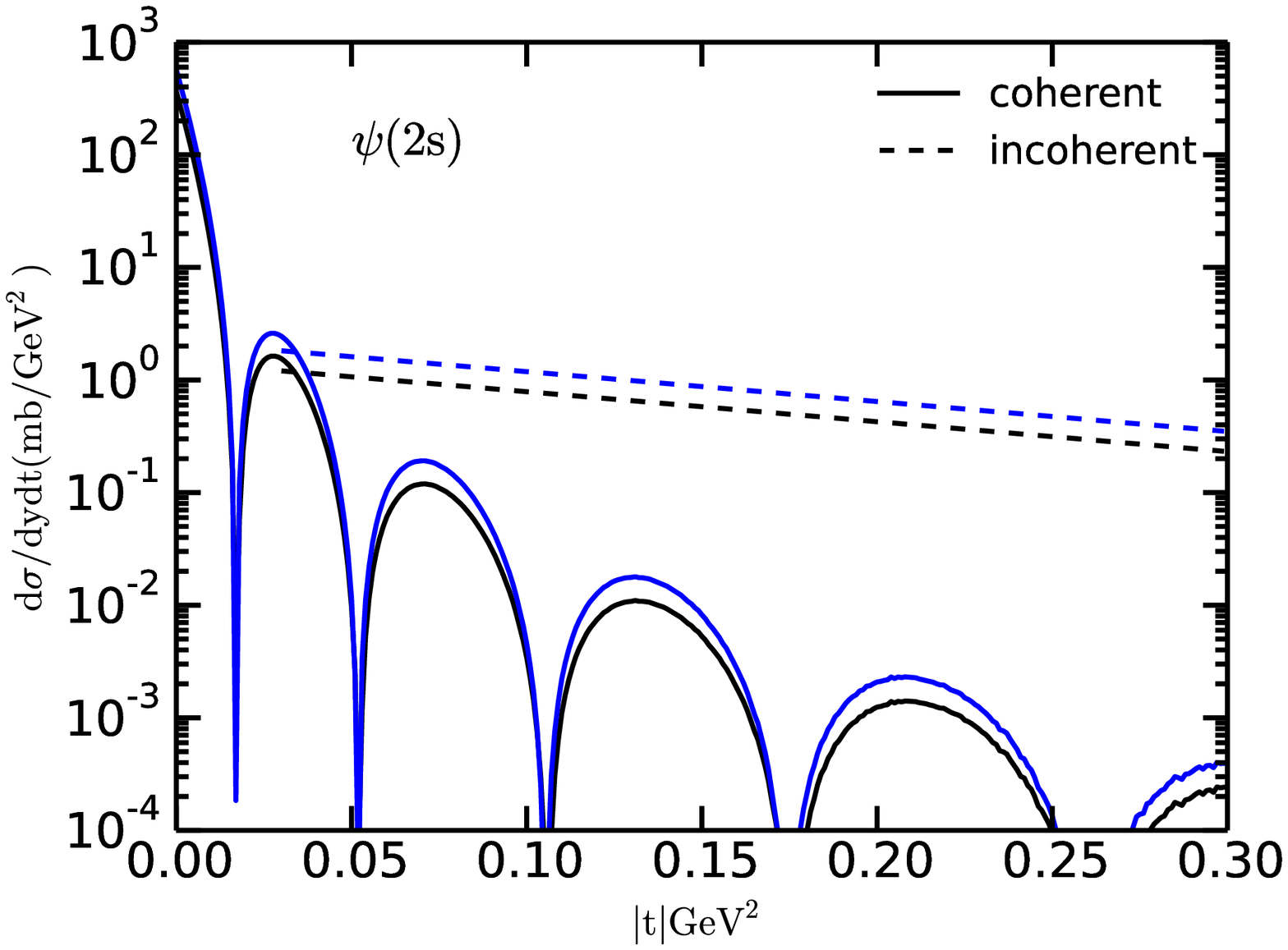}
    \caption{(Color online) Differential cross section for the coherent and incoherent production of $J/\psi$ (left panel) and $\psi(2s)$ (right panel) mesons as a function of $|t|$ at rapidity y=0 using the IIM model in PbPb collisions at $\sqrt{s_{NN}}=2.76~\text{TeV}$ (black lines) and $\sqrt{s_{NN}}=5.02~\text{TeV}$ (blue lines). The vector meson function is the Boosted Gaussian wave function.}
    \label{fig01}
    \end{figure} 
    \begin{figure}[htbp]
    \centering
       \includegraphics[width=3in]{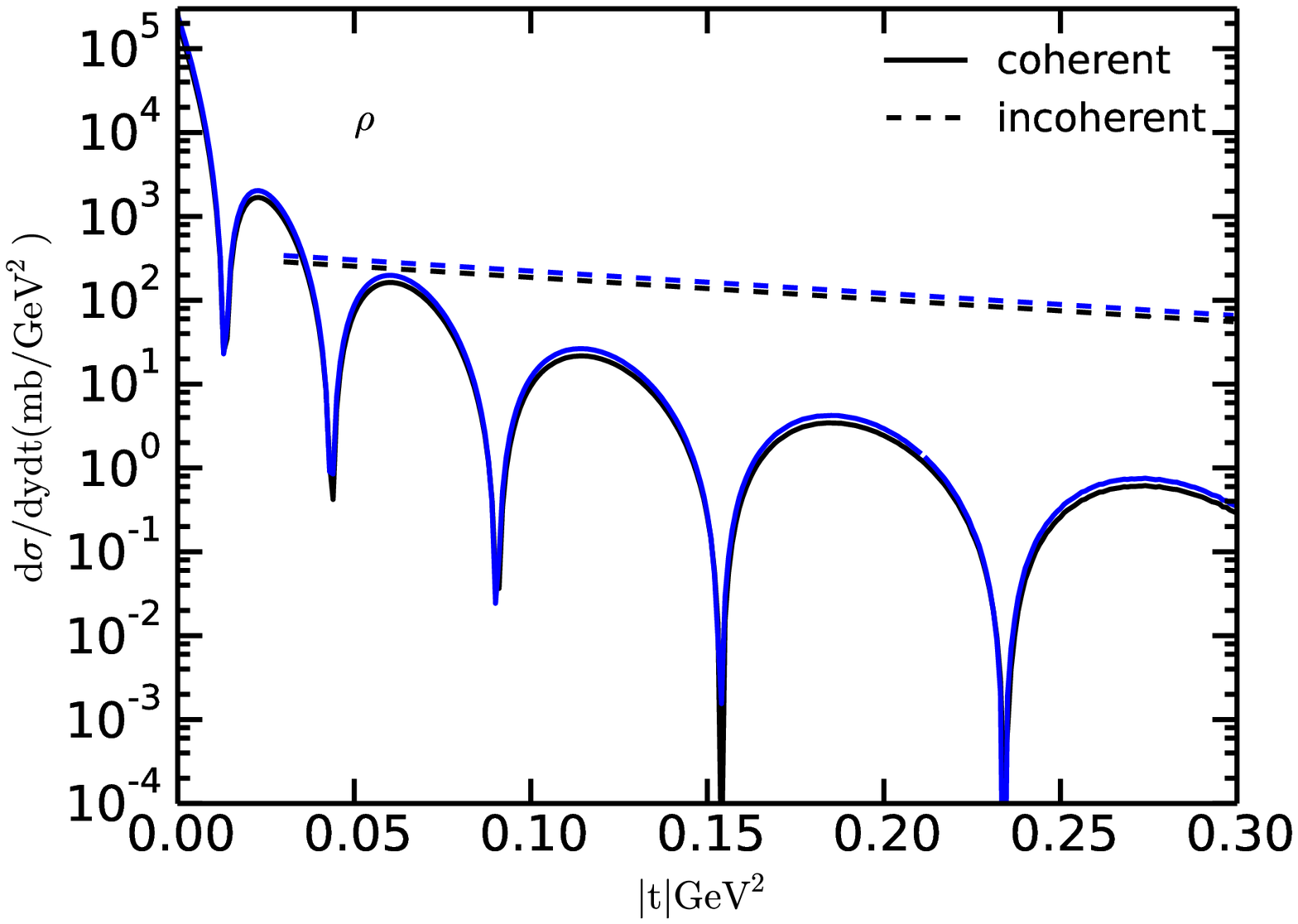}
       \includegraphics[width=3in]{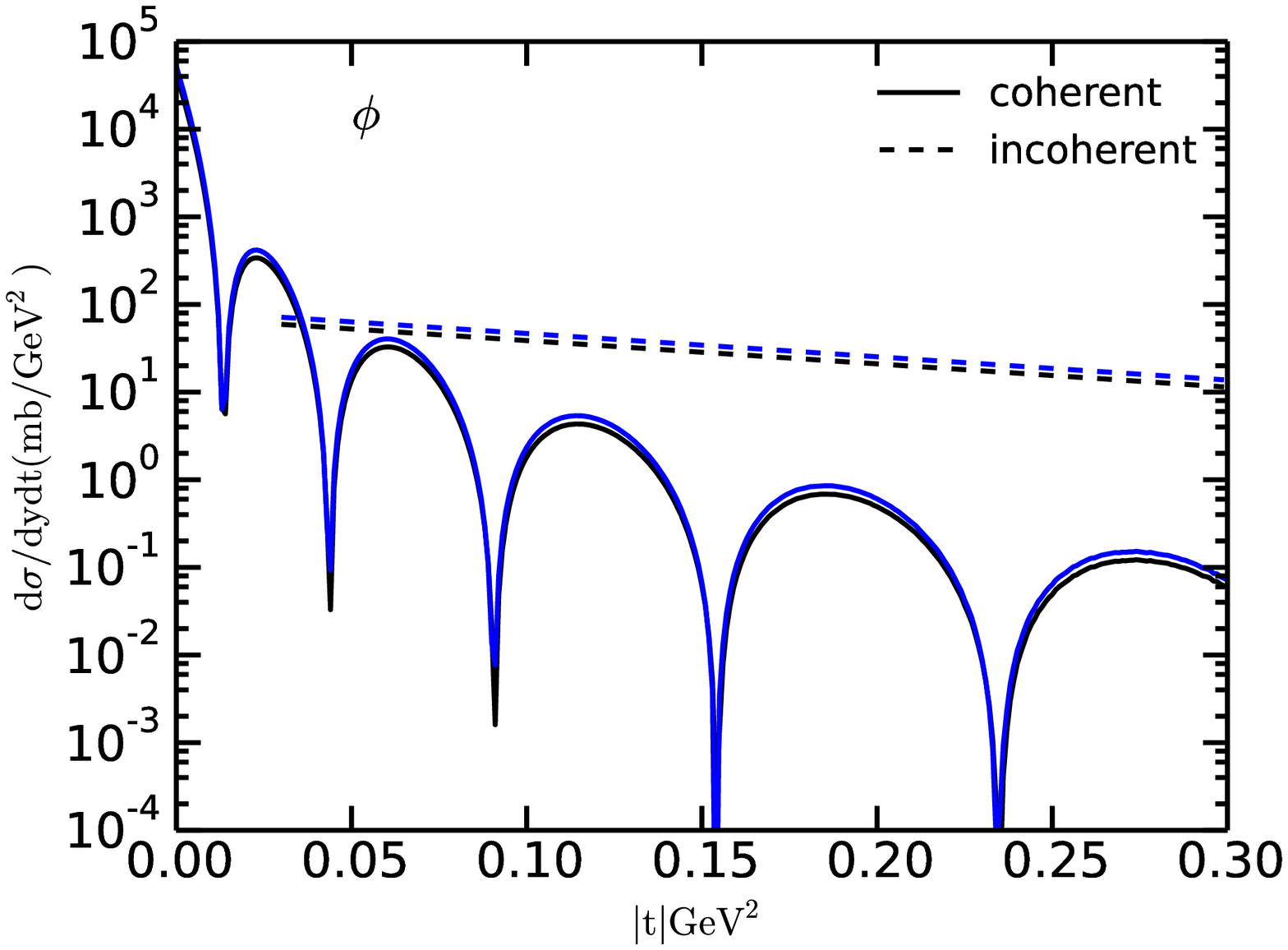}
    \caption{(Color online) Differential cross section for the coherent and incoherent production of $\rho$ (left panel) and $\phi$ (right panel) as a function of $|t|$ at rapidity y=0 using the IIM model in PbPb collisions at $\sqrt{s_{NN}}=2.76~\text{TeV}$ (black lines) and $\sqrt{s_{NN}}=5.02~\text{TeV}$ (blue lines). The vector meson function is the Boosted Gaussian wave function.}
    \label{fig02}
    \end{figure} 
    \indent At first, we calculate the coherent and incoherent differential cross section as a function of $|t|$ at mid-rapidity y=0. The differential cross section is shown in Fig.~\ref{fig01} and Fig.~\ref{fig02}. The left panels are the differential cross section of $J/\psi$ and $\rho$ mesons. The right panels are differential cross section of $\psi(2s)$ and $\phi$ mesons. The black lines are predictions at $\sqrt{s_\text{NN}}=2.76~\text{TeV}$ and the blue lines are predictions at $\sqrt{s_\text{NN}}=5.02~\text{TeV}$. It can be seen that the coherent differential cross section is very large at small $|t|$, but it decreases very fast. When $|t|$ is not small, the dominant contribution of cross section is the incoherent cross section. A similar result of the differential cross section can be found in Ref.~\cite{Goncalves:2015poa} for $\rho$ meson.\\
 \indent Secondly, we compute the diffractive rapidity distributions of $J/\psi$ and $\psi(2s)$ in the IIM model.  The rapidity distributions of $J/\psi$ and $\psi(2s)$ are computed with charm quark mass $m_c=1.27$~GeV. They are shown in Fig.~\ref{fig03}. The upper panels are predictions at $\sqrt{s_\text{NN}}=2.76~\text{TeV}$ and the lower panels are predictions at $\sqrt{s_\text{NN}}=5.02~\text{TeV}$. The left panels of Fig.~\ref{fig03} are the incoherent rapidity distributions of $J/\psi$, the black solid lines are the incoherent rapidity distributions using the Boosted Gaussian wave function,
   \begin{figure}[htbp]
   \centering
    \includegraphics[width=3in]{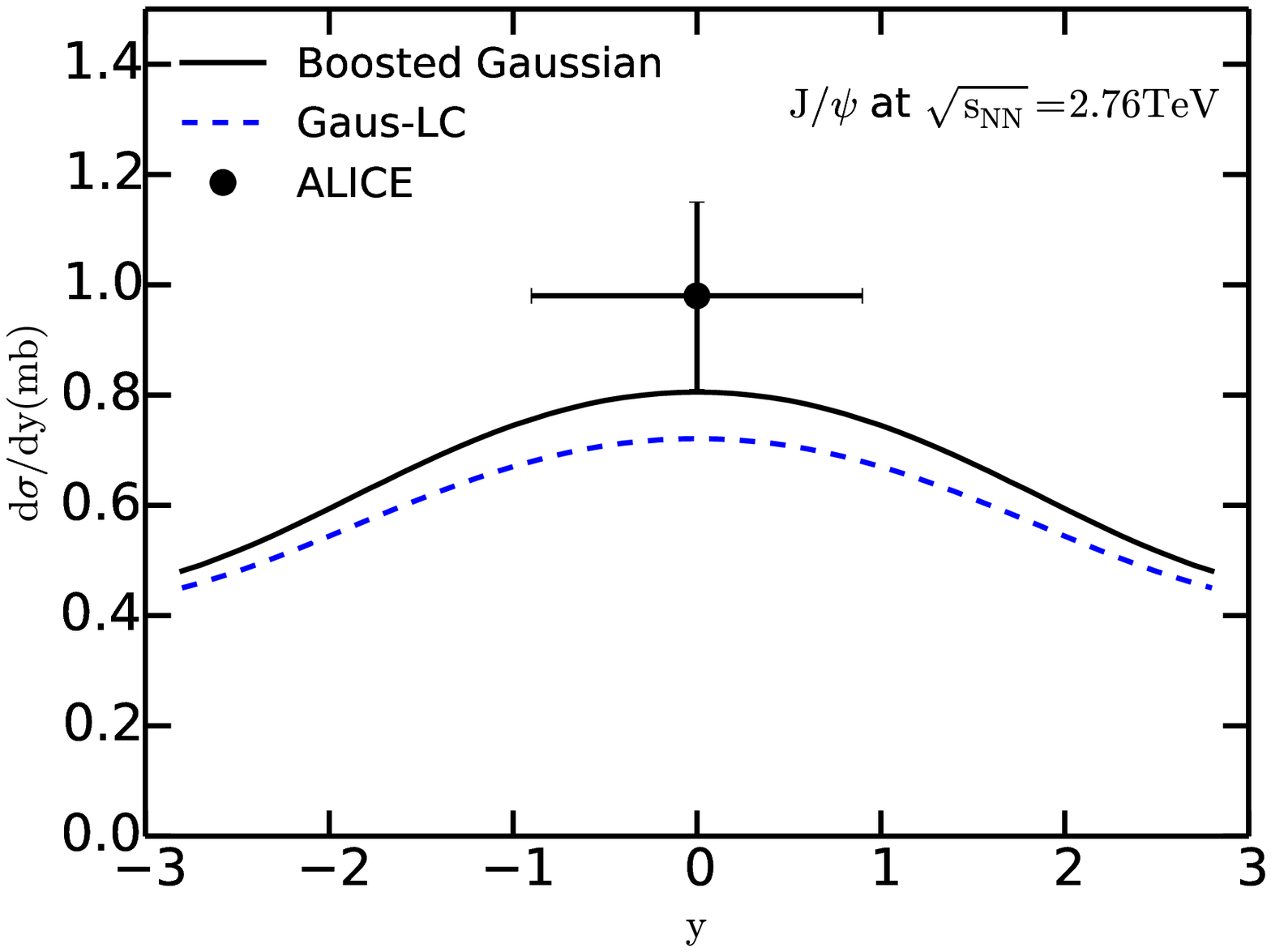}
        \includegraphics[width=3in]{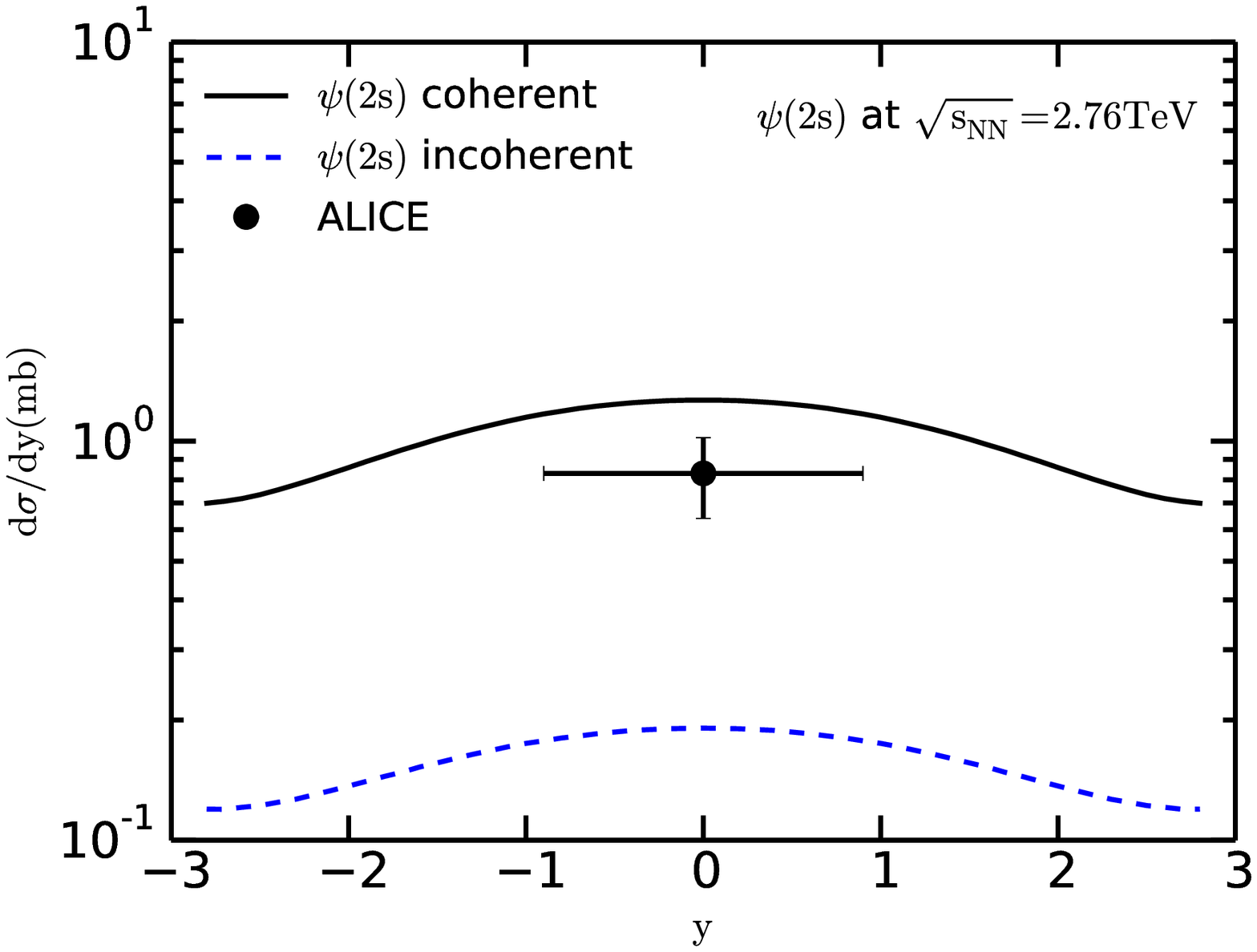}
            \includegraphics[width=3in]{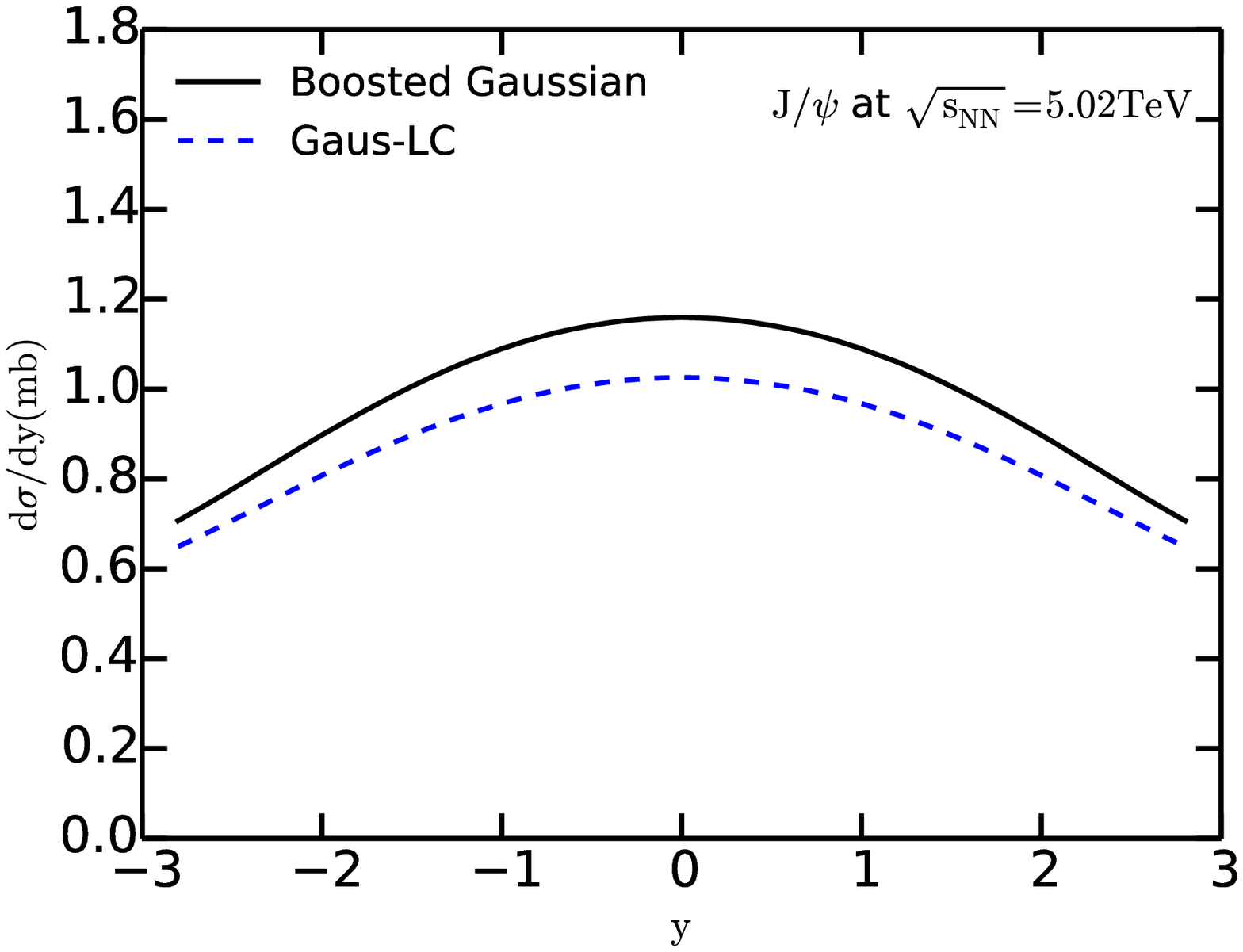}
                \includegraphics[width=3in]{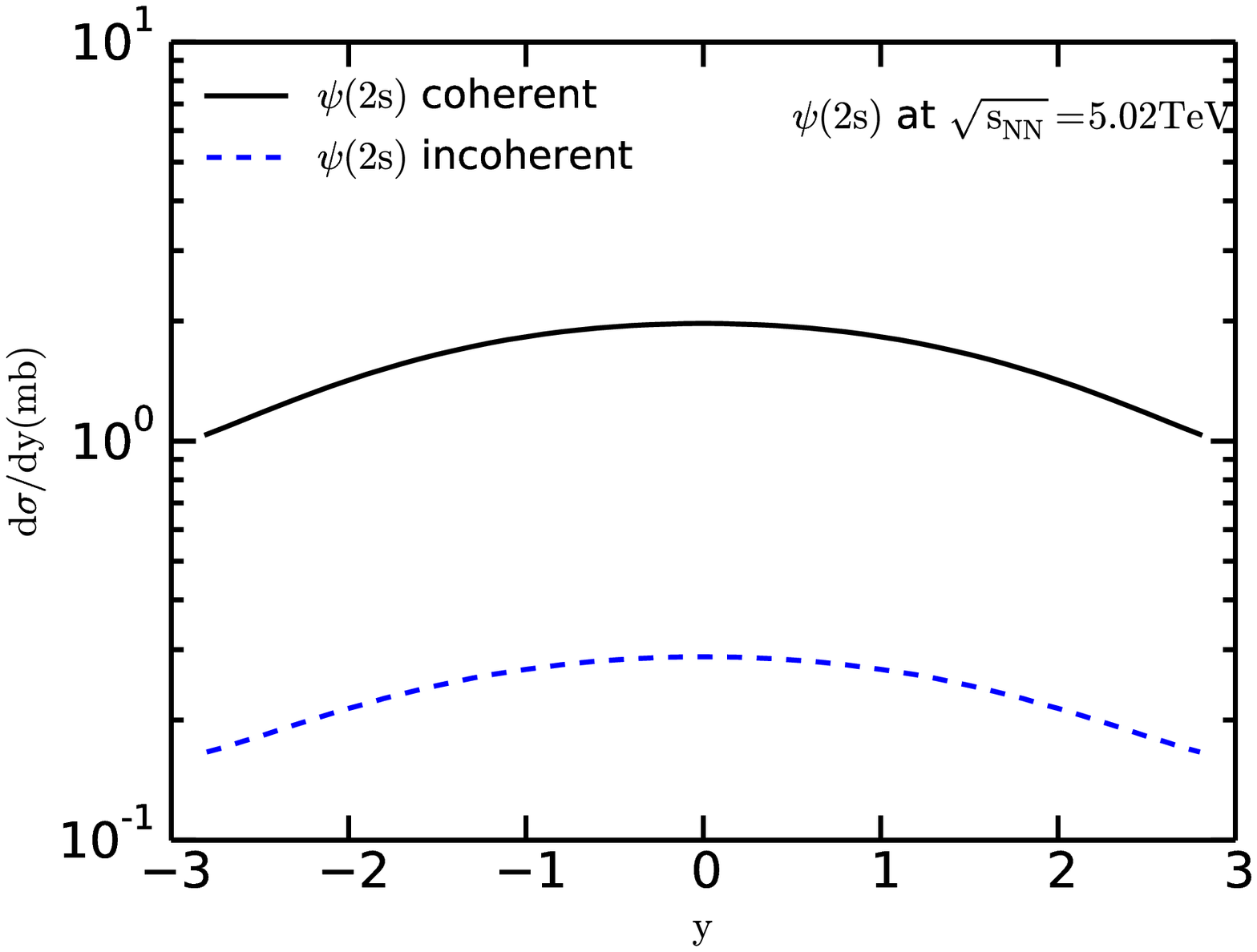}
   \caption{(Color online) Predictions for the incoherent rapidity distributions of  $J/\psi$ (left panels) calculated in the IIM model in PbPb UPCs at $\sqrt{s_\text{NN}}=2.76~\text{TeV}$ (upper panels) and $\sqrt{s_\text{NN}}=5.02~\text{TeV}$ (lower panels) using the Boosted Gaussian (black solid line) and Gaus-LC (blue dashed line) wave function. The coherent and incoherent rapidity distributions of $\psi(2s)$ (right panels) calculated in the IIM model using the Boosted Gaussian wave function. The experimental data is taken from the ALICE collaboration~\cite{Abbas:2013oua,Adam:2015sia}.}
   \label{fig03}
   \end{figure} 
 and the blue dashed lines are rapidity distribution using the Gaus-LC wave function. It can be seen that the experimental data favors the results using the Boosted Gaussian wave function from the upper left panel. In Ref.~\cite{Lappi:2013am}, the predictions of the incoherent rapidity distributions of $J/\psi$ are also computed in previous IIM parameters with $m_c=1.4$~GeV. The rapidity distribution of $J/\psi$ at mid-rapidity obtained in this work is larger than previous one using previous IIM parameters. It is because the difference of charm quark mass. 
    \begin{figure}[htbp]
    \centering
    \includegraphics[width=3in]{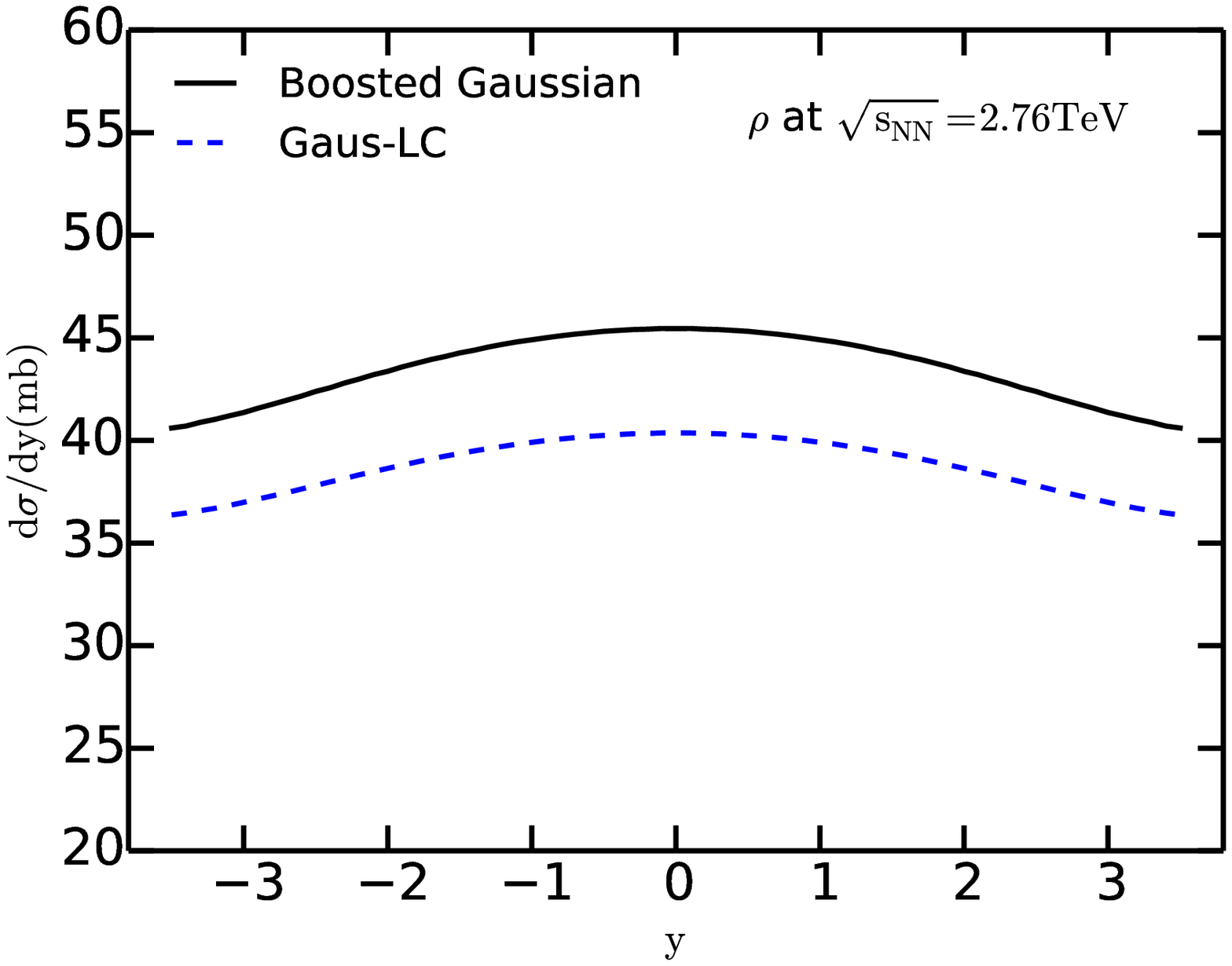}
   \includegraphics[width=3in]{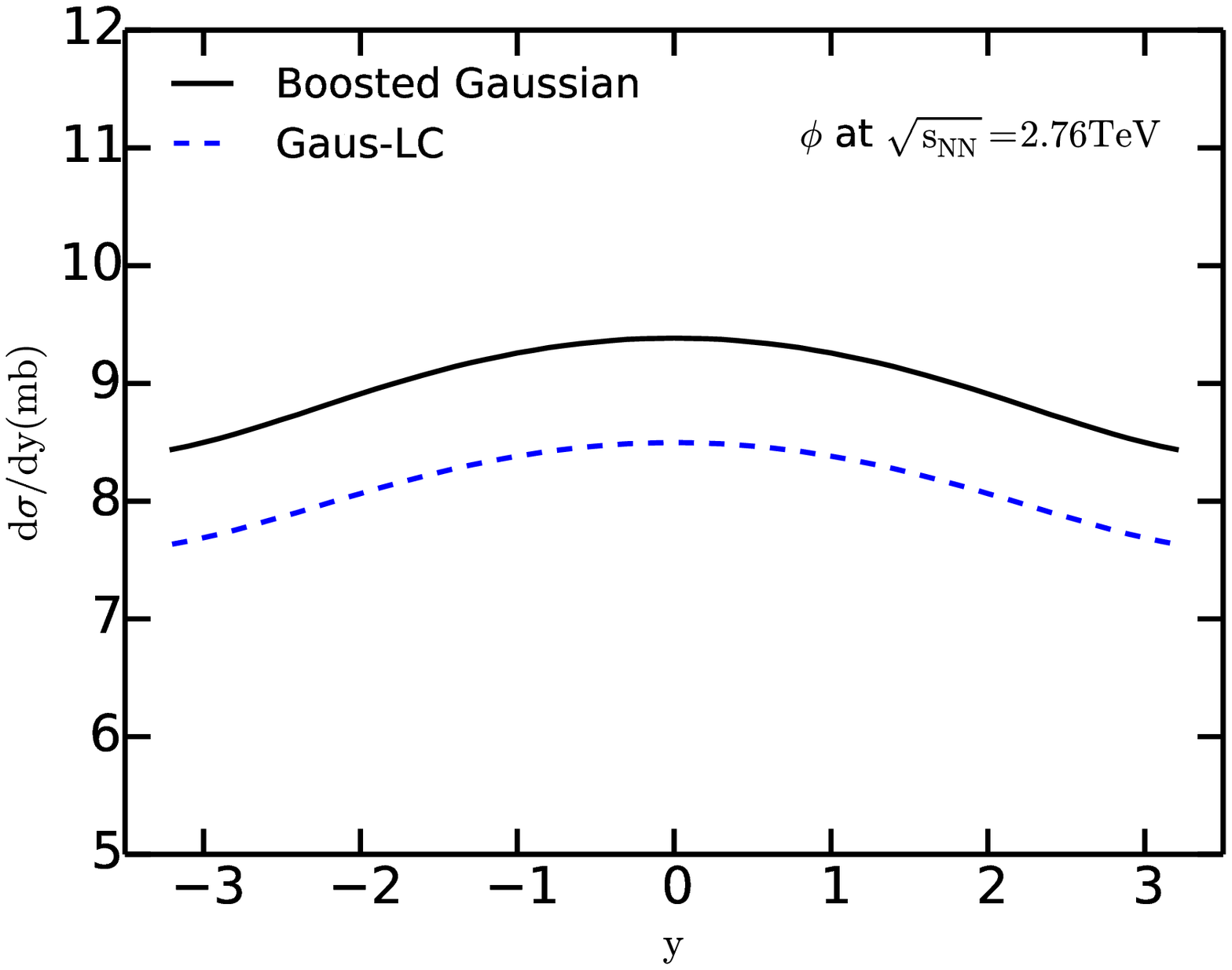}
       \includegraphics[width=3in]{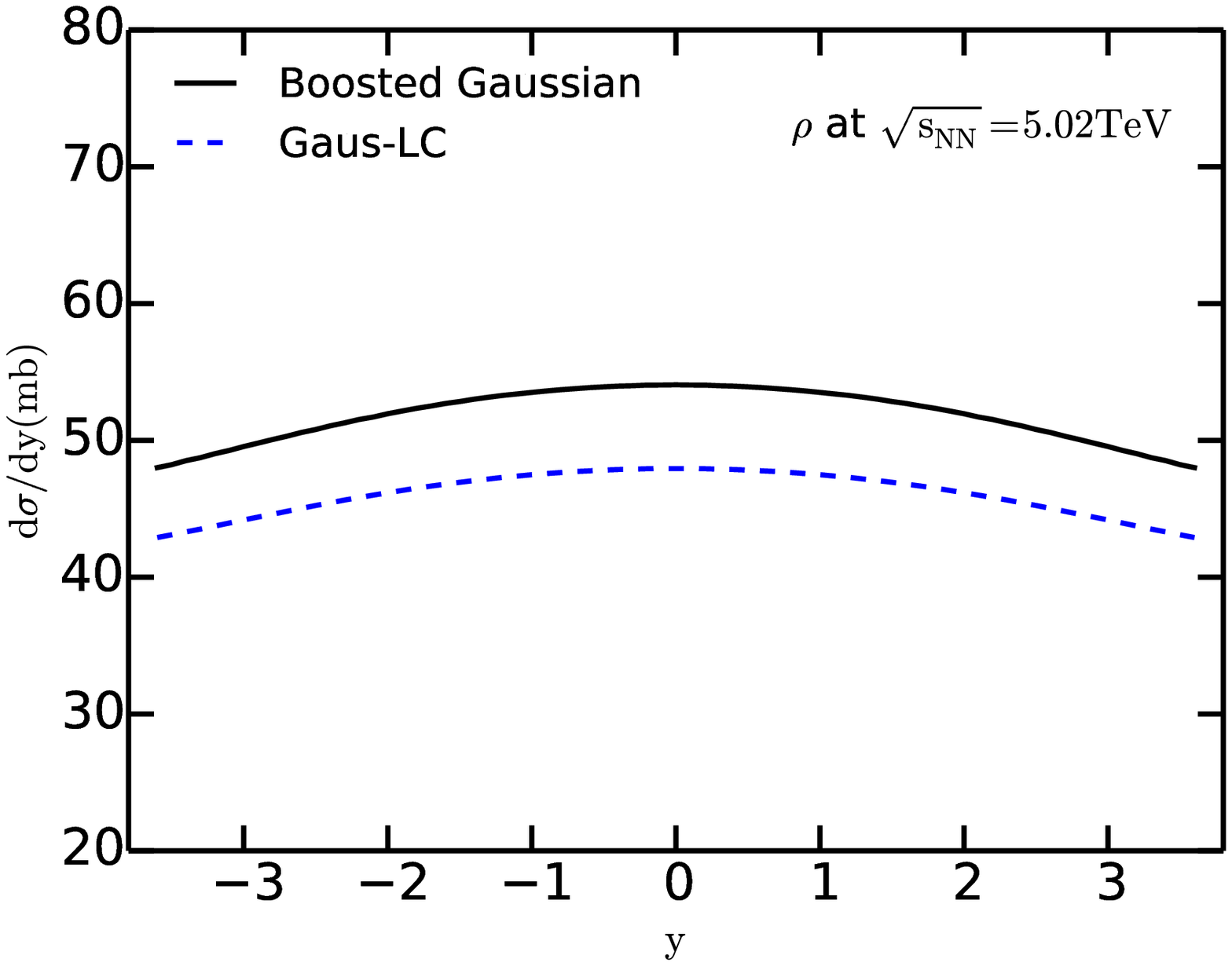}
      \includegraphics[width=3in]{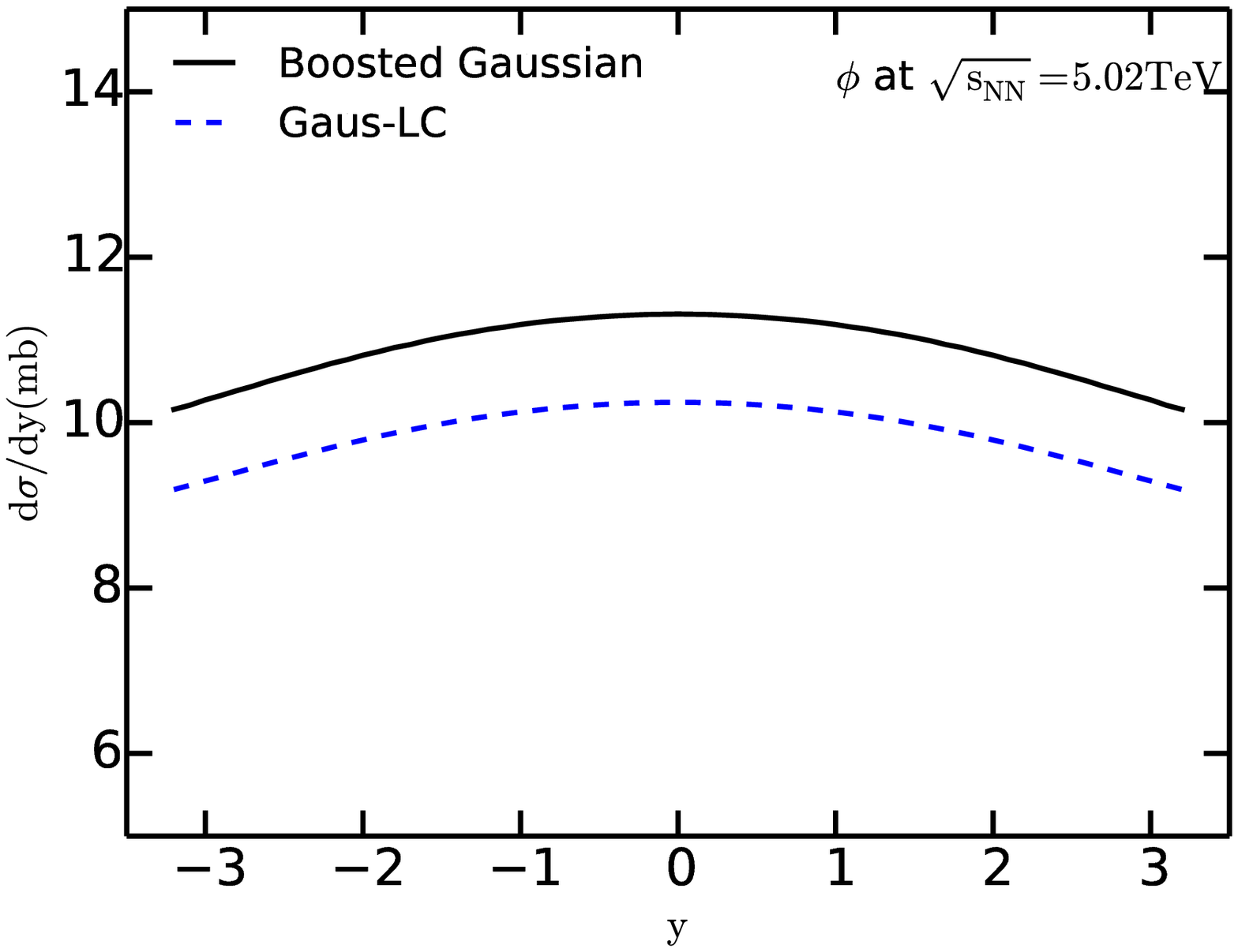}
    \caption{(Color online) Predictions for the incoherent rapidity distributions of $\rho$ (left panels) 
     and $\phi$ (right panels) calculated in the IIM model in PbPb UPCs at $\sqrt{s_\text{NN}}=2.76~\text{TeV}$ (upper panels) and $\sqrt{s_\text{NN}}=5.02~\text{TeV}$ (lower panels) using the Boosted Gaussian (black solid line) and Gaus-LC (blue dashed line) wave functions.}
    \label{fig04}
     \end{figure}
  The incoherent rapidity distribution of $J/\psi$ in PbPb UPCs at mid-rapidity measured by the ALICE collaboration is $d\sigma^\text{incoh}/d\text{y}(\text{y}=0)=0.98^{+0.19}_{-0.17}$~mb~\cite{Abbas:2013oua}. In our calculation, the prediction of the incoherent rapidity distribution for $J/\psi$ at mid-rapidity in the IIM model using the Boosted Gaussian is $d\sigma^\text{incoh}/d\text{y}(\text{y}=0)=0.80~\text{mb}$, it is still lower than the ALICE experimental data. But our prediction is closer than the prediction using old IIM model in Ref.~\cite{Lappi:2013am}. The coherent and incoherent rapidity distribution of $\psi(2s)$ is in Fig.\ref{fig03}.  The upper right panel in Fig.~\ref{fig03} are the coherent and incoherent rapidity distributions of $\psi(2s)$ at $\sqrt{s_\text{NN}}=2.76~\text{TeV}$. The black solid line is the coherent rapidity distributions using the Boosted Gaussian wave function, and the blue dashed line is the incoherent rapidity distributions using the Boosted Gaussian wave function. The prediction of  $\psi(2s)$ coherent rapidity distribution at mid-rapidity at $\sqrt{s_{NN}}=2.76~\text{TeV}$ is $d\sigma^{coh}/d\text{y}(\text{y}=0)=1.26$~mb from our calculation. The authors of Ref.~\cite{Lappi:2014eia} give prediction of  $\psi(2s)$ coherent rapidity distributions $d\sigma^\text{coh}/d\text{y}$(\text{y}=0)=0.65~mb in the IIM model using the Boosted Gaussian wave function. The experimental data measured by the ALICE collaboration is $d\sigma^\text{coh}/d\text{y}(\text{y}=0)=0.83\pm0.19$~mb~\cite{Adam:2015sia}. We also show the predictions for $\psi(2s)$ incoherent rapidity distributions in Fig.~\ref{fig03}. In our calculations, the $\psi(2s)$ incoherent rapidity distribution is expected to be $d\sigma^\text{incoh}/dy(\text{y}=0)=0.19 $~mb. In Ref.~\cite{Ducati:2013bya}, the authors predict the $\psi(2s)$ incoherent rapidity distribution $d\sigma^\text{incoh}/d\text{y}(\text{y}=0)=0.16$ mb in different formulas using the Boosted Gaussian wave function, whose parameters are not given in that paper. It can be seen that the above two predictions of incoherent rapidity distribution at mid-rapidity are close to each other. In our calculation, the wave function of $\psi(2s)$ we used is the Boosted Gaussian model. There are various sets of parameter for $\psi(2s)$ meson. Different parameter sets will give different predictions. It needs further studies for the wave function of the excited states.  \\
  \indent Finally, the incoherent rapidity distributions of $\rho$ and $\phi$ mesons in PbPb UPCs at the LHC are also calculated in the IIM model using two kinds of wave functions. They are shown in Fig.~\ref{fig04}. The black solid lines and the blue dashed lines are predictions of rapidity distributions using the Boosted Gauusain wave function and Gaus-LC wave function, respectively. The upper panels are predictions at $\sqrt{s_\text{NN}}=2.76~\text{TeV}$ and the lower panels are predictions at $\sqrt{s_\text{NN}}=5.02~\text{TeV}$.
  Since there is no experimental data for  $\rho$ and $\phi$ mesons incoherent rapidity distributions for now, we can only compare our results with other theoretical predictions. Considering the uncertainty of the incoherent rapidity distributions of $J/\psi$, it is expected that $d\sigma^{\text{incoh}}/d\text{y}(\text{y}=0)=50\pm10$~mb and $d\sigma^{\text{incoh}}/d\text{y}(\text{y}=0)=10\pm2$~mb for $\rho$ and $\phi$ in PbPb UPCs at $\sqrt{s_{NN}}=2.76~TeV$ from our calculations. The predictions of the incoherent rapidity distributions for $\rho$ meson at PbPb UPCs has been computed using other approach in Ref.~\cite{Santos:2014vwa}, where the rapidity distribution at mid-rapidity is expected  $d\sigma^{\text{incoh}}/d\text{y}(\text{y}=0)=30\pm10$~mb at $\sqrt{s_\text{NN}}=~2.76~\text{TeV}$. It can be seen that our predictions are close to the previous predictions. \\
\section{conclusion}
 \indent In this paper, we have evaluated the vector mesons incoherent rapidity distributions in PbPb UPCs at $\sqrt{s_\text{NN}}=2.76~\text{TeV}$ and $\sqrt{s_\text{NN}}=5.02~\text{TeV}$. The rapidity distributions of $J/\psi$, $\psi(2s)$, $\rho$ and $\phi$ mesons are computed in the IIM model, whose parameters are fitted from the combined ZEUS and H1 data released in 2015. The incoherent rapidity distributions for $J/\psi$ meson has been measured by the ALICE collaboration. The predictions of the IIM model using the Boosted Gaussian wave function are closer to the experimental data of the ALICE collaboration than the previous predictions. We  have also predicted the coherent and incoherent rapidity distributions of $\psi(2s)$ in the IIM model using the Boosted Gaussian wave function. We compare our results with previous predictions of $\psi(2s)$. The rapidity distributions for $rho$ and $phi$ mesons are also evaluated in the IIM model using two kinds different vector meson functions at the LHC. The experimental data of the incoherent rapidity distributions of $\psi(2s)$, $\rho$ and $\phi$ in PbPb UPCs at the LHC are absent now. We hope the experimental data of $\psi(2s)$, $\rho$, $\phi$ will be measured in the future and thereby we can compare them with the theoretical predictions. 

\section{Acknowledgements}
We thank the useful discussions with V. P. Gon\c{c}alves, Bo-Wen~Xiao, M.~Ahmady and H. ~M\"antysaari. This work is supported in part by the National 973 project in China (No:~2014CB845406).

\end{document}